\documentclass[11pt]{article}

\usepackage[utf8]{inputenc}
\usepackage[T1]{fontenc}
\usepackage{helvet}
\usepackage{setspace}
\usepackage{verbatim, authblk}
\usepackage{ulem}
\usepackage{graphicx} 
\usepackage{textcomp}
\usepackage{array}
\usepackage{makecell}
\usepackage[a4paper, top=2cm,bottom=2cm,left=1in,right=1in,marginparwidth=1.75cm]{geometry}
\usepackage{xcolor}
\usepackage{amsmath}
\usepackage{float}
\usepackage{stmaryrd}
\usepackage{comment}

\usepackage[square,numbers]{natbib}

    \newcommand{\ex}{\mbox{$\underline{e_x}$}}  
    \newcommand{\ey}{\mbox{$\underline{e_y}$}}  
    \newcommand{\ez}{\mbox{$\underline{e_z}$}}  

\title{Spanwise dispersion optimizes the efficiency of dense microfluidic trap arrays}

\author[,1]{Nicolas Ruyssen\thanks{Corresponding author: \texttt{nicolas.ruyssen@sorbonne-universite.fr}}}
\author[2, 3]{Gabriel Fina}
\author[1]{Rachele Allena}
\author[4]{Marie-Caroline Jullien}
\author[2,3]{Jacques Fattaccioli}

\affil[1]{Arts et Métiers Institute of Technology, Université Paris 13, Sorbonne Paris Cite, IBHGC, HESAM Universite, 75013 Paris, France}
\affil[2]{PASTEUR, Département de Chimie, École Normale Supérieure, PSL Université, Sorbonne Université, CNRS, 75005 Paris, France}
\affil[3]{Institut Pierre-Gilles de Gennes pour la Microfluidique, 75005 Paris, France}
\affil[4]{Univ. Rennes, CNRS, IPR (Institut de Physique de Rennes) UMR $6251$, F-$35000$ Rennes}

\date{}

\begin{document}
\raggedbottom
\maketitle


\bigskip

\begin{abstract}
Microfluidic Trap Arrays (MTAs) have proved efficient tools for several applications
requiring working at the single cell level like cancer understanding and treatment or immune
synapse research. Unfortunately, it generally appears that many traps stay empty, even after
a long time of injection which can drastically reduce the number of samples available for post-treatment. It has been shown that these unfilled traps were due to the symmetrical nature of the flow around the traps, with a break in symmetry improving capture efficiency. In this work, we use a numerical approach to show that it is possible to generate optimal geometries that significantly improve capture efficiency. This efficiency is associated with an increase in the lateral dispersion of the objects; we show that adding disorder to the layout of the traps is the most optimal solution and may stay very efficient independently of the trap array size. These numerical results are corroborated by experiments, validating our approach.
    
\end{abstract}

\section{Introduction}
Among the last discoveries in medicine and biology research, Microfluidic Trap Arrays (MTAs) appear as very promising tools for several applications like tumorous cells response to drugs \cite{Wlodkowic2010,Dereli-Korkut2014}, leukemia cells identification \cite{Lee2008}, stem cells agregates analysis \cite{Jackson2017} or immune response triggering \cite{Pinon2022a,Pineau2022a}. This success is due to the fact that MTAs allow an accurate spatio-temporal control of numerous and various isolated micro-sized objects such as cells \cite{DiCarlo2006,Pinon2022a,Skelley2009,Wlodkowic2009,Challier2021}, droplets \cite{Pompano2011,Carreras2017,Bai2010,Mesdjian2021,Pinon2022a,Fradet2011}, or beads \cite{SohrabiKashani2019, Mesdjian2021, Challier2021}. Hydrodynamic MTAs working principle is the following: objects are diluted inside a fluid going through a microfluidic chamber constituted by an array of a large number of traps. While following the fluid flow by viscous drag force, objects can be caught in the traps individually or by small groups. Trapping can be achieved passively by gravity \cite{Charnley2009,Fradet2011, Figueroa2010, Rousset2017}, inertial lift \cite{Rousset2017}, permanent magnets \cite{Winkleman2004,Smistrup2006} or actively using valves \cite{Zhou2016,Au2011}, optical tweezers \cite{Grigorenko2008}, dielectrophoresis \cite{Rosenthal2005,Challier2021}, electromagnets \cite{Lee2004} and acoustics \cite{Evander2007}. However, these techniques are contingent on the physical properties of the objects, frequently necessitating more intricate designs, materials, manufacturing processes, and incurring higher costs, thereby constraining their integration into regular clinical practice. On the contrary, single layer hydrodynamic MTAs are now easy to fabricate with standard soft lithography process, their mechanism depends only on the carrier fluid properties and objects geometry. Nonetheless, many traps can stay empty even after a long time of injection which can drastically reduce the number of exploitable objects and raises the important question of hydrodynamic trapping efficiency optimization \cite{Kobel2010,Mesdjian2021}. In order to improve capture performance, a numerical approach seems perfectly appropriate for exploring different experimental configurations (geometry, flow properties, density of objects to be captured, etc...) To this end, we can cite \cite{Kobel2010,Deng2014,Jin2015,Xu2013_mef,Xu2013_exp,Wang2021,Rousset2017} who improved the trap and channels geometry or \cite{Kim2012,SohrabiKashani2019} who enhanced the trap relative positions and orientations with respect to the main flow. However, all these numerical studies focused on MTAs with a few number of traps. Recently, we experimentally validated a Computational Fluid Dynamics (CFD) and particle tracing approach in a rectangular MTA of $100$ traps \cite{Mesdjian2021}. Interestingly, we proved the importance of the flow structure in the chamber and of breaking symmetries along the cavity to enhance the capture efficient by favouring spanwise dispersion. In this work, we propose to go a deep further by adapting our CFD/particle model to a fully parameterized geometry and to couple it with optimization algorithms to maximize the MTA's trapping efficiency. In a first approach, the trapping efficiency is optimized independently on five dimensionless geometric parameters. In a second set of numerical experiments, we show that similar trapping efficiencies can be reached using disordered trap arrays. The physical interpretation we propose is based on considering the flow structure along the cavity. The best geometry that is numerically generated is tested experimentally and the filling efficiency is experimentally recovered. We believe this new geometry opens new routes for very large scale trapping of biological objects and could contribute to the generation of patient specific biological big data at the single cell scale.

\section{Materials and methods}
\label{section_MM}

The whole numerical model is developed using the commercial Finite Element software Comsol Multiphysics \textregistered \ while post-processing is done using Python \textregistered  . 

\subsection{Parametrized geometry}
\begin{figure}[H]
    \centering
    \includegraphics[width = \textwidth]{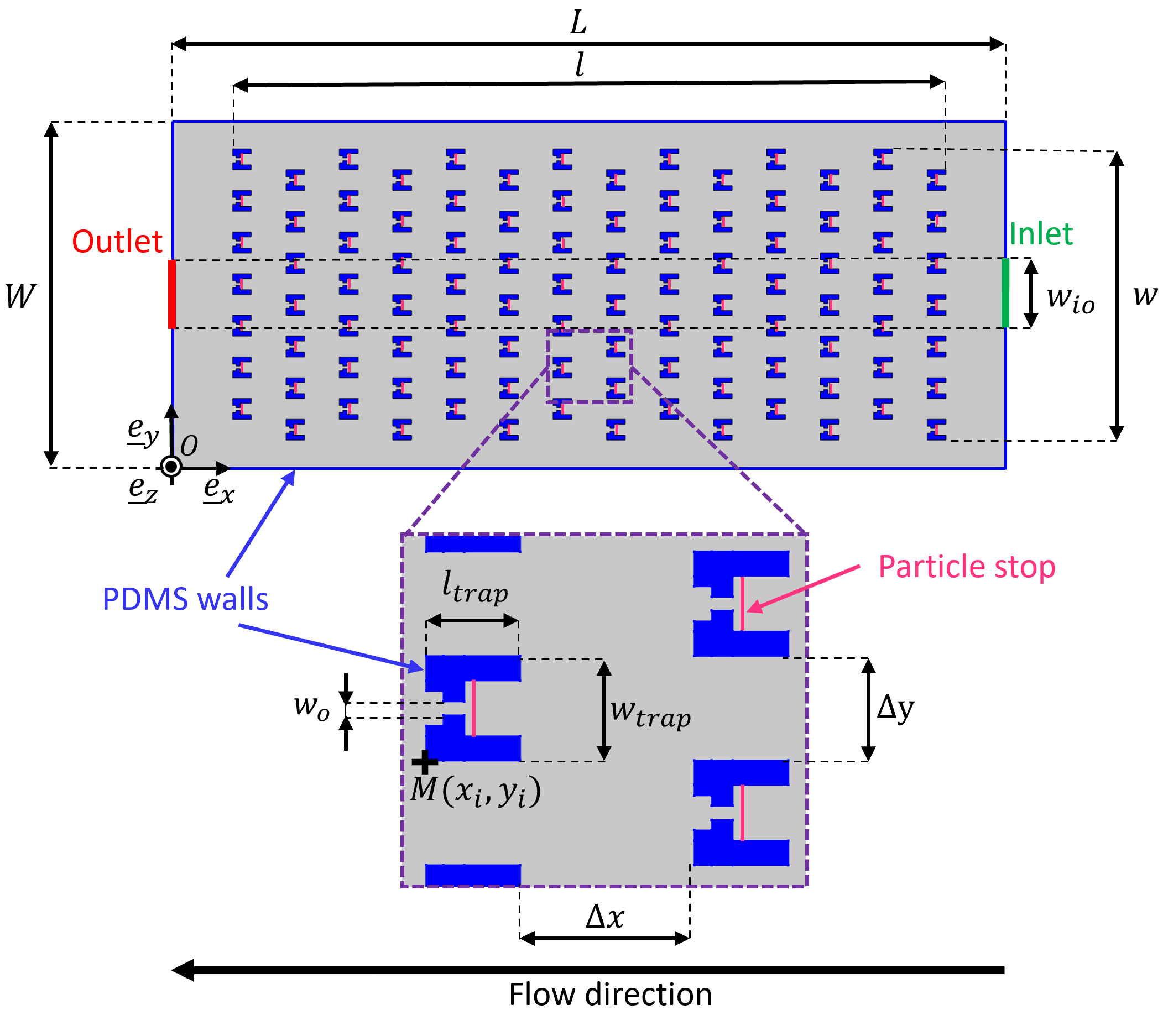}
    \caption{The model geometry. The fluid domain is represented in grey, the traps in turquoise, the inlet resp outlet channels in green resp in red and the PDMS walls in blue. A particle stop frontier is added in each trap (in pink) to virtually stop particles without modifying the fluid flow.}
    \label{fig_geom}
\end{figure}

We consider a single layer MTA of $98$ traps as in \cite{Mesdjian2021}. Because of symmetry, a $2$D model built from parametrized geometric primitives is suggested (Figure \ref{fig_geom}). The MTA is initially organized in a rectangular staggered pattern of axis $(Ox)$ in the streamwise direction and of axis $(Oy)$ in the spanwise direction, (see Figure \ref{fig_geom} where all the geometrical parameters are sketched). The dimensions of the MTA are $l$ and $w$ and the one of the cavity are $L$ and $W$ in the streamwise resp spanwise directions ; and the distances between the traps are respectively $\Delta x$ and $\Delta y$. We call "chamber" the whole system including the MTA and the cavity. For more information, all the fixed geometric parameters of the model are reported Table \ref{tableau_all_param} section \ref{section_SI}. 

\subsection{Physics of the problem}

We consider the flow as incompressible, stationary and the fluid as Newtonian. Furthermore, the maximal Reynolds number $Re=\frac{Uh}{\nu}$ (where $U \sim 1$ mm s$^{-1}$ is the maximal mean velocity, $h = 14$ $\mu  $m the thickness of the cavity and $\nu = 10^{-6}$ m$^2$ s$^{-1}$ the kinematic viscosity of the solution) is about $Re = 0,01 $, which allow us to consider the Stokes equations. In order to simplify the study by considering a 2D flow in the in-plan of the cavity (\textit{ie} $(O,\ex, \ey)$, Figure \ref{fig_geom}), we consider the Brinkman equations that account for both the viscous dissipation in the out-of-plane of the cavity (\textit{ie} $(O,\ez)$, Figure \ref{fig_geom}) through a force $\underline{f}$, and the viscous dissipation in the in-plane. As such, the Brinkman equations write : 
\begin{equation}
\begin{cases}
    - \nabla p + \mu \ \nabla^2 \underline{v}+ \underline{f}= \underline{0} \\
    \hfill \nabla.\underline{v} = 0 
\end{cases}
    \label{eq_stokes_incomp}
\end{equation}
where $p$ is the pressure field, $\mu$ the fluid dynamic viscosity, $\underline{v}$ the velocity field. The chamber is horizontally orientated such that gravity does not contribute to the flow. The friction force $\underline{f}$ is given in the shallow channel approximation: 
\begin{equation}
    \underline{f}= - \frac{12 \ \mu}{h^2} \ \underline{v} 
    \label{eq_ssa}
\end{equation}
Therefore, $\underline{v}$ corresponds to the depth-averaged velocity field in-plane (\textit{ie} $(O,\ex, \ey)$, Figure \ref{fig_geom}) to reduce the flow in $2$D. As boundary conditions, uniform pressures are applied on inlet $p_{in} =100 $ Pa and outlet $p_{out} =0 $ Pa boundaries. On the PDMS walls, the viscous fluid-solid interaction forces a zero fluid velocity $\underline{v} = \underline{0}$. \\
Experimentally, we study a dispersion of polystyrene (PS) particles of diameter $D = 5 \ \mu$m and of concentration $ C_p = 10^6$ mL$^{-1}$. This concentration leads to a typical distance between particles about $62$ $\mu$m which allows neglecting particle interactions. Furthermore, their diffusion coefficient is about 8.7 10$^{-14}$ m$^2$.s$^{-1}$ which can be considered negligible compared to the convection of the particles and the concentration is considered sufficiently small that they do not modify the viscosity of the suspending fluid. Despite a low Reynolds number, the Lifto-Diffusif number is ${\cal N} = \frac{27\pi V^2\rho R_p^4}{2k_b T h} \sim 30$, thus, lift forces are non-negligible in our flow conditions \cite{Mottin2021Nov}. However, the corresponding maximal depletion thickness at the outlet of the chamber is $\delta = \sqrt{\frac{R_p \rho L }{W h^2 \mu}} \sim 1,3 \ \mu $m which is small compared the particle size of $5$ $\mu$m, such deviations will thus be disregarded in the following. The PS particles density of $1,05$ g mL$^{-1}$ is very close to the carrier fluid one, thus particle sedimentation will be also disregarded. As the Reynolds number is low and that the particles are smaller than the channel height, we make the assumption that the particle trajectories are similar to the carrier fluid streamlines. Their equation of motion is obtained considering a purely advective transport :
\begin{equation}
    \frac{\partial \underline{r} }{\partial t} = \underline{v} 
    \label{PFD_point}
\end{equation}
where $\underline{r}$ is the particle positions. When a particle enters into a trap, it is artificially stopped at a stop boundary in each trap (pink boundaries Figure \ref{fig_geom}). $ N_p = 200 \approx 2 \ N_{traps}$ particles are uniformly located on the inlet boundary (green boundary Figure \ref{fig_geom}). 

\subsection{Optimization of a staggered trap array}

From a purely applications perspective, the reader can refer to sections \ref{sec_resultats_opti} outlining the optimum characteristics of a high-performance MTA. This section presents the definition of the different parameters that are used for the optimization of the MTA. We first have to define the objective of the optimization. For this, we define the filling efficiency $E_{fill}=N_{occupied}/N_{traps}$ of a given geometry as the number of traps occupied by at least one particle divided by the total number of traps and the capture efficiency $E_{capt}=N_{capt}/N_{traps}$ as the number of trapped particles divided by the total number of particles. In order to optimize these objectives, several geometric parameters, defined arbitrarily by us, have been varied. We thus define five dimensionless geometric parameters of the MTA, defined as:
 
 \begin{itemize}
     \item The centering $C$ represents the shift between the inlet and outlet channels in the spanwise direction and thus allows studying symmetry/asymmetry effects between the inlet and the outlet. Its value is $100 \% $ when the inlet and the outlet are perfectly aligned (as in Figure \ref{fig_geom}) and $0$ when the shift is maximal. $C$ is given by:
     \begin{equation}
          C = 1 - \frac{\Delta y_{io}}{\Delta y_{io \ max}} = 1-\frac{2 \ \Delta y_{io}}{W-w_{io}}
     \end{equation}
     where $\Delta y_{io}$ is the spanwise distance between the inlet and outlet and $\Delta y_{io \ max}$ its maximal value.
     
     \item The width ratio  $W_r$ in the spanwise direction between the trap array's and the cavity's width defined by: 
     \begin{equation}
         W_r = \frac{w}{W}
     \end{equation}
    This ratio allows scrutinising the role of a particle bypass on the sides of the cavity.
  
     \item The length ratio $L_r$ in the streamwise direction between the trap array's and the cavity's length defined by:
     \begin{equation}
         L_r = \frac{l}{L}
     \end{equation}
      This ratio allows investigating possible entrance effect of particle distribution.

     \item The trap array aspect ratio $A_r$ defined by the ratio between the number of lines $N_l$ in streamwise direction and the number of columns $N_c$ in spanwise direction of the trap array: 
     \begin{equation}
         A_r = \frac{N_l}{N_c}
     \end{equation}
    This ratio is maybe the less intuitive, but we will see that it may play a role. Most studies show MTA with an $A_r>1$ but without motivating this choice. We thus intend to inspect its role. 
    
     \item  And finally, in order to caracterize the importance of the the channel width occupied by particles at the inlet, we introduce, the inlet and outlet channels to cavity width ratio in the spanwise direction: 
    \begin{equation}
        Wch_{r} = \frac{w_{io}}{W}
    \end{equation}
    
 \end{itemize}

The interest of working with dimensionless parameters is to have theoretically adjustable results in a similar system but at larger or smaller scale while respecting the same hypotheses on the fluid flow.

\subsection{Optimization by disorder in the trap array}
\label{sec_desordre}
All the MTAs reported in the literature concern regular trap position structures. We therefore question the role of this regularity and how the results are affected when disorder is introduced into the traps positioning. To do so, we start from a regular trap position structure  (Figure \ref{fig_geom}) and introduce irregularities in the trap network. This is achieved by applying a randomized translation along the horizontal and vertical directions of each trap "$i$":

\begin{equation}
    \begin{cases}
    x_i = x_{oi} + \psi_x(i) \ D_f \ \Delta x \\
    y_i = y_{oi} + \psi_y(i) \ D_f \ \Delta y \\
    \end{cases}
    \label{eq_random_pos}
\end{equation}
where $(x_i,y_i)$ are the coordinates of the $i^{th}$ trap bottom left corner (point $M$ in Figure \ref{fig_geom}), $(x_{0i},y_{0i})$ its original coordinates following the rectangular staggered pattern, $\psi_x$ and $\psi_y$ are independent random distributions of numbers between $-1$ and $1$, $D_f$ a positive constant called "disorder factor" between $0$ and $1$. We tested two ways of introducing disorder into the lattice, uniform and centered Gaussian with a standard deviation of $0.3$ distributions for $\psi_x$ and $\psi_y$. 

\subsection{Trapping experiments}

\begin{figure}[H]
    \centering
    \includegraphics[width=\textwidth]{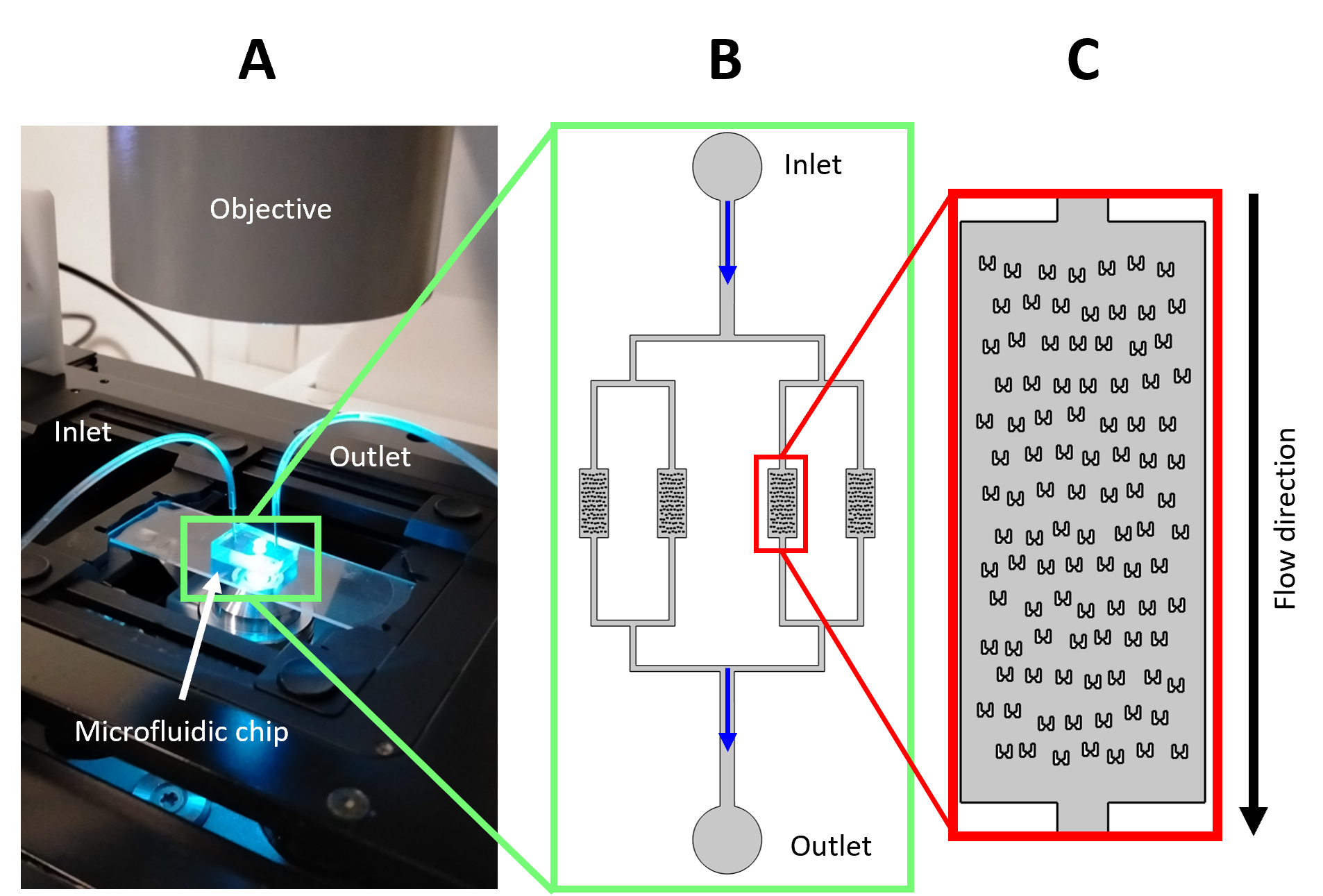}
    \caption{Experimental setup. \textbf{A}: the microfluidic chip containing the channels is placed on the microscope plate, its inlet and outlet are connected via tubes to a pressure regulator. \textbf{B}: scheme of the $4$ microfluidic chambers linked to the inlet and outlet where the tubes are anchored. \textbf{C}: scheme of a zoom on the tested disordered chamber.}
    \label{fig_protocole}
\end{figure}

The precise microfabrication steps are the same as those for \cite{Mesdjian2021} and are very briefly mentioned in this section. Microfluidic chamber resulting from disorder study is fabricated using single layer standard photo-lithography techniques. Based on previous MTA studies in the literature, we chose to connect $4$ MTAs in parallel \cite{Skelley2009} (Figure \ref{fig_protocole}B and C). A chrome photomask is fabricated from the geometry files of optimization studies by direct laser writing. A circular $h=14$ $\mu$m  thick layer of photoresist resin is generated on a silicon wafer following a spin coating protocol. A  master is obtained by UV-light activation and dissolution of the non-activated photoresist resin. PDMS is mold on the master and the microfluidic chip is closed by sticking a glass coverslip on the chip following a plasma activation protocol. The microfluidic chip is then observed using fluorescent microscopy with a 10X magnification  (Figure \ref{fig_protocole}A). The inlet and outlet channels are connected to a  pressure controler with a pressure drop $ \Delta p $ corresponding to a flow rate $Q_v = 0,5 \ \mu$L \ min$^{-1}$ and a $ C_p = 10^6$ mL$^{-1}$ concentrated solution of fluorescent particles, hydraulic resistance variation of the chip is negligible during all the experiment \cite{Mesdjian2021}.

\section{Results and discussion}
\label{section_Results}

\subsection{All parameters influence capture and filling efficiency which are correlated}
\begin{figure}[H]
    \centering
    \includegraphics[width=\textwidth]{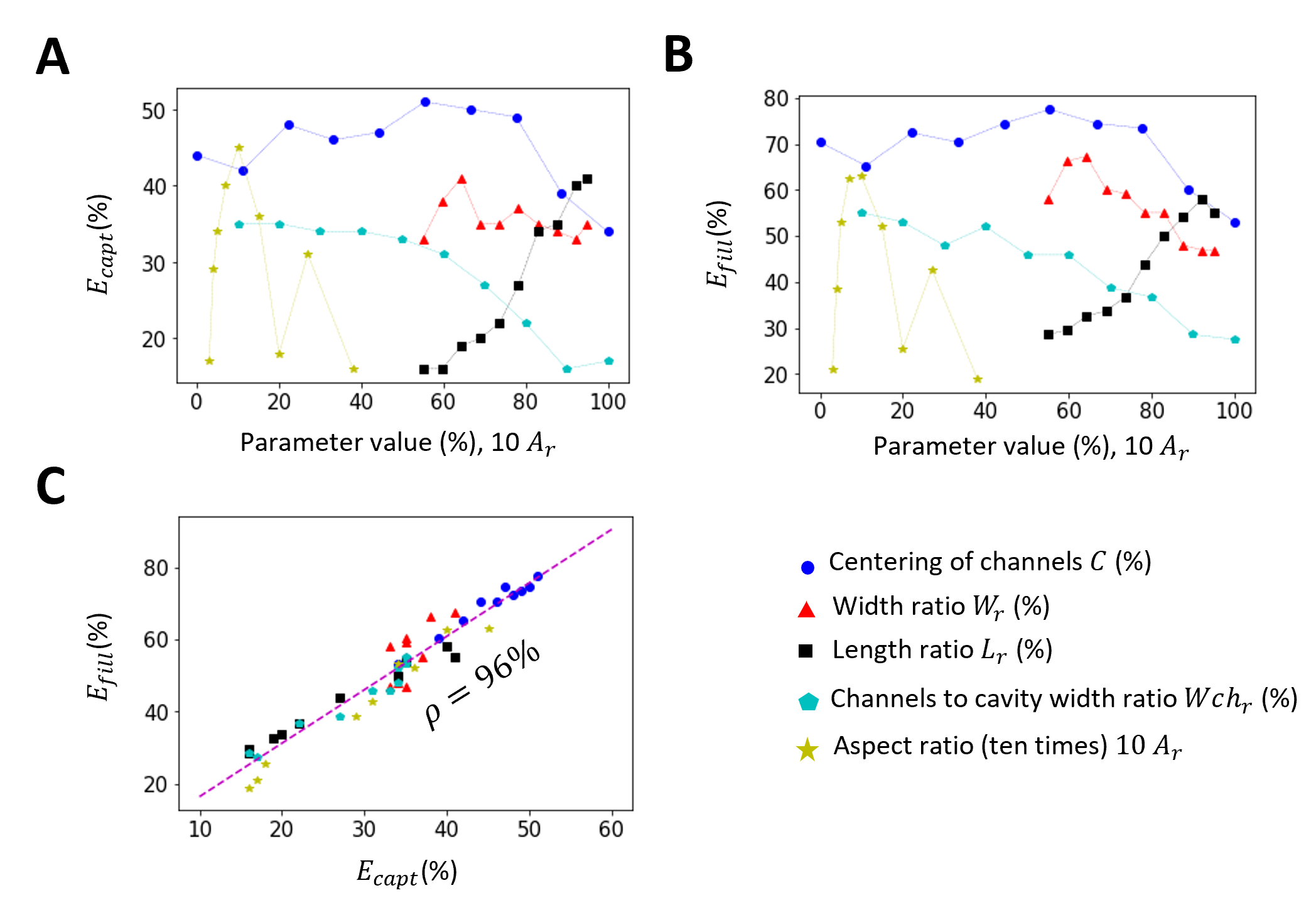}
    \caption{Influence of the five dimensionless parameters $C$, $W_r$, $L_r$, $A_r$ and $Wch_r$ on capture efficiency (\textbf{A}) and filling efficiency (\textbf{B}). When parameters are fixed, their respective values are : $C=100\%$, $W_r = 84\% $, $L_r = 86\%$, $A_r = 2$ and $Wch_r = 23\%$ which corresponds to reference geometry Figure \ref{fig_geom}. To plot every curve on the same graph, $A_r$ is multiplied by $10$ and interpolation curves are plotted to facilitate reading only. Filling efficiency is represented as function of capture efficiency in \textbf{C}, the linear correlation coefficient is $96\%$.}
    \label{Fig_influence_param}
\end{figure}

Figure \ref{Fig_influence_param} summarises the evolution of capture (A) and filling (B) efficiencies for a regular MTA for the $5$ first dimensionless parameters : $C$, $W_r$, $L_r$, $A_r$ and $Wch_r$. As such, the influence of each parameter is investigated while keeping the other parameters constant. 

\begin{figure}[H]
    \centering
    \includegraphics{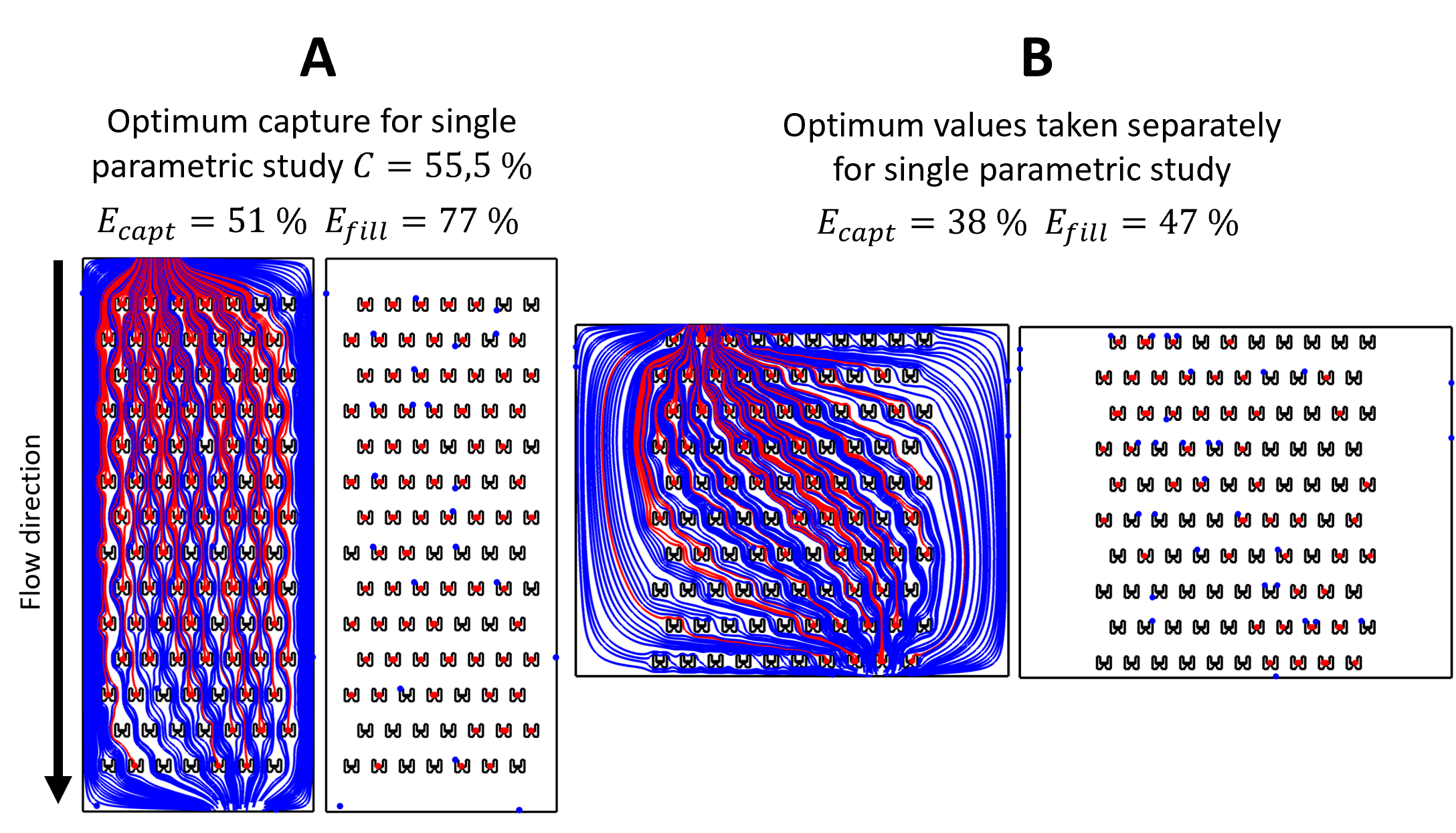}
    \caption{Particle trajectories and final positions for the best geometry resulting from single parametric study (\textbf{A}) and for the geometry obtained combining each optimum values from the same study (\textbf{B}), which appears as poorly efficient, justifying a multiparametric approach. }
    \label{fig_traj_pram_simple}
\end{figure}

At first glance, it is clear that all these parameters have an influence on $E_{fill}$ up to $ \pm 100\%$ and with  maximal and minimal values at $77\% $ and $15\% $. Setting properly these parameters is therefore crucial for a good chamber design. It is difficult to extract clear trends in capture/filling efficiencies as a function of the various parameters, as most of these trends are non-monotonic. We would nevertheless like to draw attention to the role of the centring parameter, which shows that the best capture and filling efficiencies, of  $51\% $ and $77\% $, are obtained for a relative inlet/outlet centering between $44\%$ and $66\%$ (Figure \ref{Fig_influence_param}A). This is the best configuration so far. In addition, we notice that this geometry recovers an oblique flow (Figure \ref{fig_traj_pram_simple}) which looks like an improved version of our previously presented oblique chamber \cite{Mesdjian2021}. Figure \ref{Fig_influence_param}C shows the correlation between the two measures $E_{fill}$ and $E_{capt}$ which is close to a linear correlation coefficient of $96\%$. Therefore, the optimum parameters values for both efficiencies are very close and optimizing the MTA geometry with $E_{fill}$ or $E_{capt}$ as objective functions should give similar trends. 

The strong variations between points do not allow intuitive explanations about the influence of the parameters on capture/filling efficiencies and suggest interdependence of parameters on them. Thus, a simple optimization approach fixing each parameters to their optimal values from this first study do not necessarily result in an efficient geometry. Indeed, the geometry built with each optimum values taken separately exhibits poor capture ($38\%$) and filling efficiencies ($47\%$) with respect to the optimal one, see Figure  \ref{fig_traj_pram_simple}B. We deduce that optimization of MTA-chamber trapping efficiency is a complex and non-intuitive process and is about finding a compromise between combinations of geometric parameters which justifies a multiparametric approach.

\subsection{Optimization of a staggered trap array leads to a large oblique flow}
\label{sec_resultats_opti}
\begin{figure}[H]
    \centering
    \includegraphics[width=\textwidth]{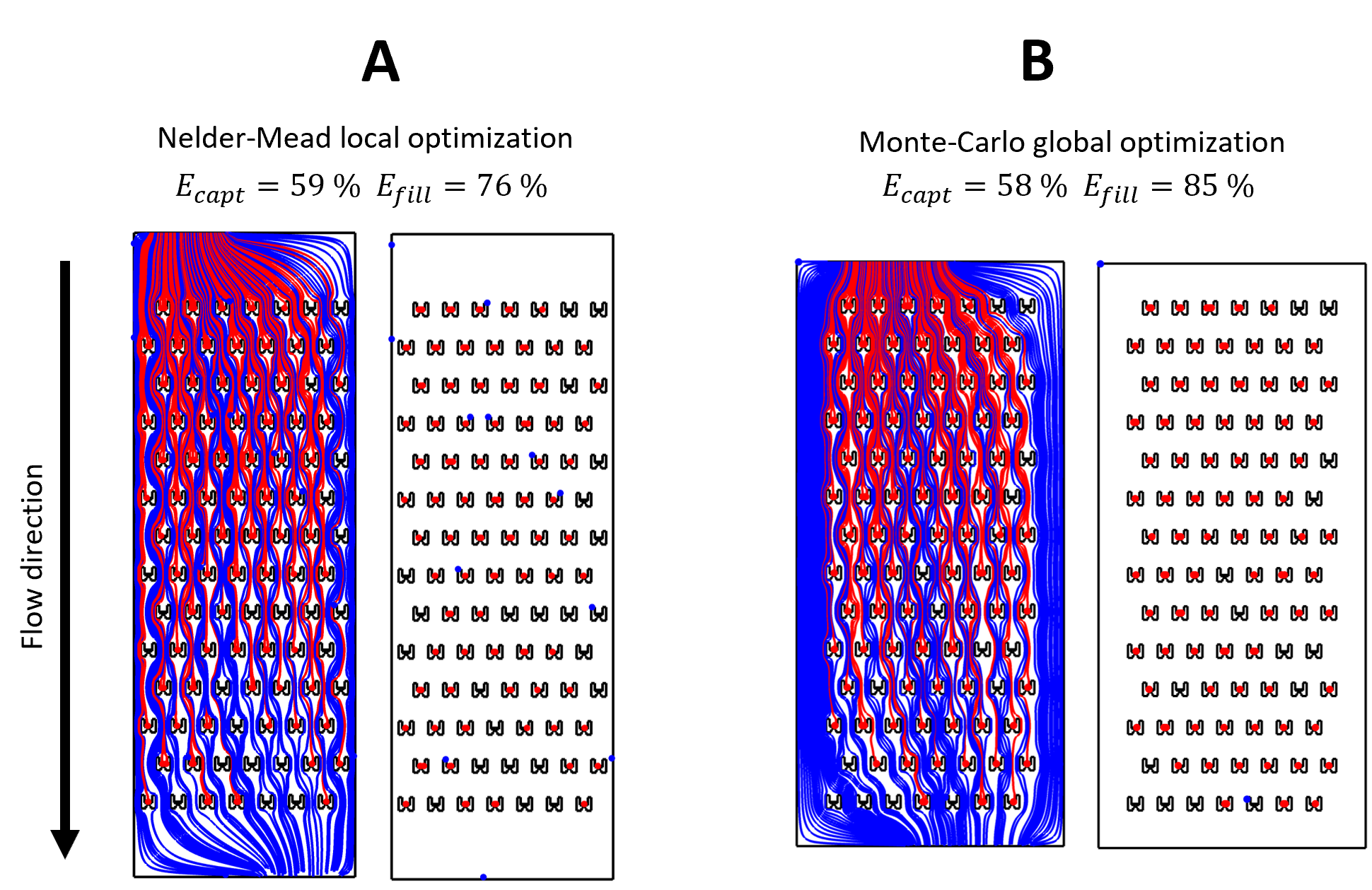}
    \caption{Particle trajectories and final positions in the locally optimized chamber (\textbf{A}) and in the globally optimized chamber (\textbf{B}). The trapped particles are represented in red, the others in blue.}
    \label{traj_opti}
\end{figure}

In this section, the MTA is let as staggered and several optimization algorithms are coupled with the CFD/particles model. The optimization problem consists in finding optimal set of parameter values leading to a maximum (local or global) of an objective function, chosen here as capture efficiency : $E_{capt}(C, W_r,L_r,A_r,Wch_r)$. Local optimization finds a local maximum of $E_{capt}$ depending on the initial set of parameter values while global optimization look for the global maximum (if exists) of $E_{capt}$ \cite{methodes_opti}. We expect the global optimization to provide the best geometry because this optimization is supposed to test a larger number of configurations and becomes independent of the initial parameter values that are implemented. We performed local optimization for different initial geometries with the simplex based Nelder-Mead algorithm \cite{NelderMead}. All these studies lead to different optimized geometries, which confirms the presence of several local maxima of $E_{capt}$. This presence of local maxima lead us to use global optimization techniques using the Monte Carlo algorithm with a number of model evaluations limited to $2000$ \cite{ComsolDoc}. \\
Figure \ref{traj_opti} shows the particles trajectories and final positions for the locally (A) and globally (B) optimized geometries. The best local optimization is obtained for the geometry having the following parameter values: $C = 19\% $, $W_r = 93\% $, $L_r= 78\%$, $A_r= 50\%$, $Wch_r = 37\%$ and the global one: $C= 44\% $, $W_r=78\% $, $L_r = 87\%$, $A_r= 50\%$, $Wch_r= 48\%$. Despite finding different parameter values, their capture efficiencies are very close of respective values $59\%$ and $58\%$, but with a significant difference in filling efficiency ($76\%$ and $85\%$) in favor of the globally optimized chamber. Interestingly, despite the fact that the two geometries are not identical, it seems difficult to reach a filling efficiency better than about $85\%$. In addition, the fact that $C \neq 1$ confirms the importance of an oblique flow that allows increasing capture and filling efficiency. We believe that the symmetry breaking between the upstream and downstream flow allows a spanwise dispersion of the particles. As such, mass transport in the spanwise direction of the flow should be improved by adding disorder in the design of the MTA which is the subject of the next section.

\subsection{Disordered trap arrays drastically increase capture \& filling efficiency} 

\begin{figure}[H]
    \centering
    \includegraphics[width =\textwidth]{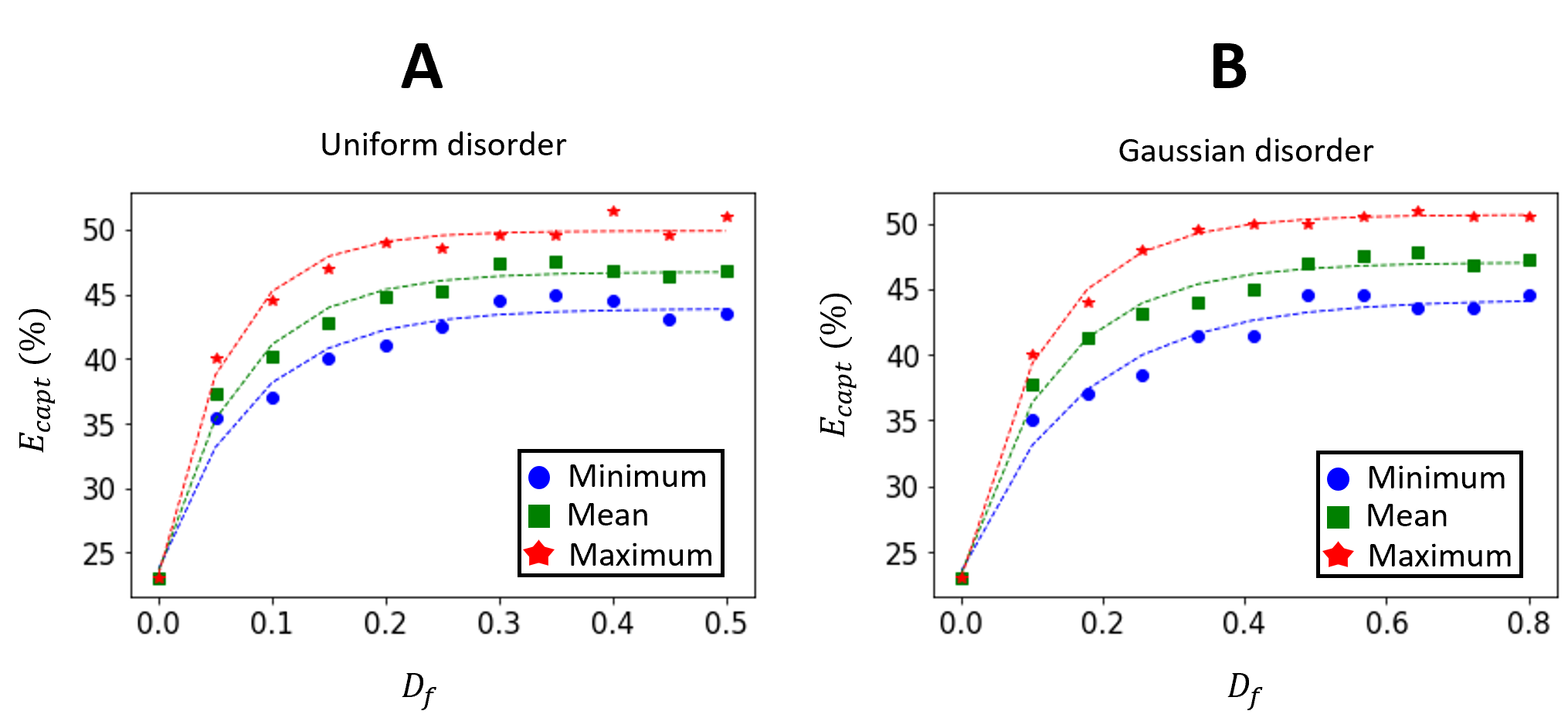}
    \caption{Evolution of capture efficiency with respect to a uniform distribution (\textbf{A}) and a Gaussian distribution (\textbf{B}) of the disorder factor $D_f$. For each $D_f$ value, the geometry is re-built $10$ times. Maximum, Minimum and Mean curves correspond respectively to best, worst and average $E_{capt}$ for the $10$ repeated geometries.}
    \label{Fig_graph_desordre}

\end{figure}

In this section we investigate how the addition of disorder on the trap positioning may affect capture and filling efficiencies. For this purpose, we explore geometries generated as described in section \ref{sec_desordre}. The initial geometry is a symmetric cavity and channels (shown Figure \ref{fig_geom}) of poor capture efficiency  ($E_{capt} = 23\%$) as the chamber has an inlet/outlet symmetry. Disorder is then added in the MTA using $10$ different values of the disorder factor $D_f$ introduced in section \ref{sec_desordre}. As the MTA geometry is built with probability laws, the repeatability of the results is assessed by rebuilding the geometry $10$ times for each $D_f$ value. 
Figure \ref{Fig_graph_desordre} shows the evolution of $E_{capt}$ with respect to  $D_f$ for uniform (A) and Gaussian distributions (B) of the disorder factor $D_f$. We notice that for a same $D_f$ value, a variety of different geometries can be obtained, leading to a dispersity of capture efficiencies of the order of $ 10\%$ (bounded by the maximum and minimum values). For both optimization methods, we observe that the capture efficiency increases with the disorder parameter and reaches a plateau about $50\% $ at best, around $D_f\sim 0.3$. A local maxima of capture efficiencies is identified to be of $52\%$ and $51\%$ for both respectively uniform and Gaussian distributions which is slightly less as for the optimization of staggered MTA. When the capture efficiency plateau is reached, the best geometries are respectively obtained for $D_f= 0,40 $ and $D_f = 0,64 $ for uniform and Gaussian distributions respectively. The simulation data are in good agreement with a first order exponential law : 

\begin{equation}
    E_{capt} = E_{0} + \Delta E_{capt}\left(1-\exp\left(\frac{D_f}{D_f^*}\right)\right)
\end{equation}
where $E_0 = 23\% $ is the capture efficiency without disorder, $\Delta E_{capt}$ represents the maximum possible gain of capture efficiency and $D_f^*$ a characteristic disorder factor corresponding to one third of the minimum disorder factor needed to obtain $95\%$ of the capture efficiency gain. For the mean curves, the corresponding $D_f^*$ are respectively $0,08$ (uniform), $0,12$ (Gaussian) and $\Delta E_{capt} = 23\%$ for both distributions which means that the capture efficiency of a chamber with symmetric cavity and channels is doubled by the disordered trap pattern.
  
\begin{figure}[H]
    \centering
    \includegraphics[width = \textwidth]{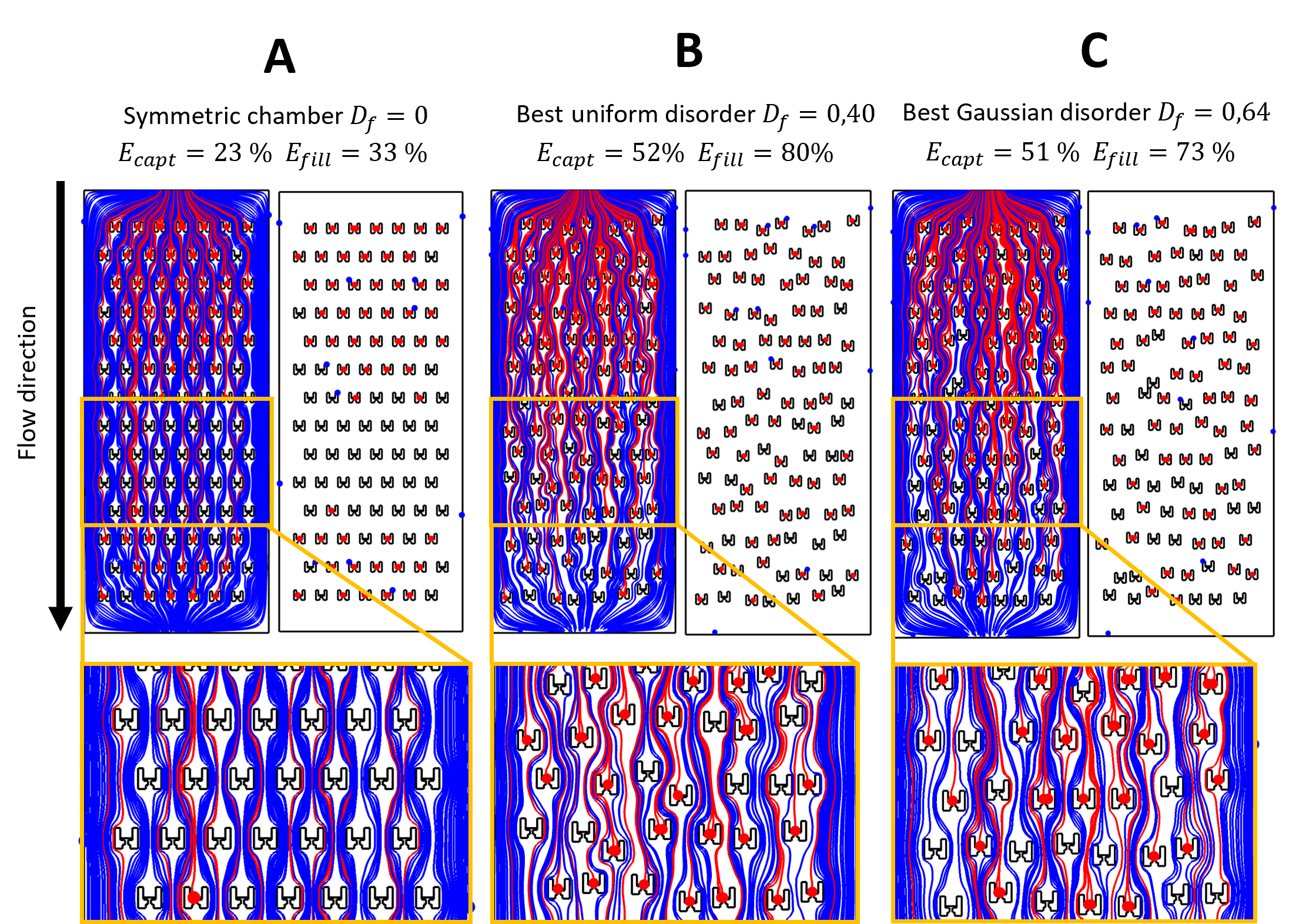}
    \caption{Particle trajectories and final positions in a symmetric chamber (\textbf{A}), the  disordered chamber for the best uniform (\textbf{B}) and Gaussian (\textbf{C}) distributions. Trapped particles are represented in red, the others in blue.}
    \label{Fig_traj_desordre}
\end{figure}

\noindent Figure \ref{Fig_traj_desordre} shows the particle trajectories and final positions for the symmetric chamber (A), the best uniform (B) and best Gaussian (C) distributions of the disorder factor. In order to be able to identify trapped particles, these are identified on entry in red, while those that are not trapped are indicated in blue. The respective capture and filling efficiencies are $ 23\% $, $ 33\% $ (symmetric chamber), $ 52\% $, $ 80\% $ (uniform distribution of $D_f$) and $ 51\% $, $ 73\% $ (Gaussian distribution of $D_f$). The slight difference in filling efficiency is in favor of the uniform distribution  of $D_f$. Moreover, this performance is independent from the size of the trap array in the spanwise direction when using inlet and outlet channels as wide as the cavity (supplementary Figure ).

In a symmetric chamber where inlet and outlet channels are perfectly aligned, streamlines exhibit upstream/downstream symmetry along the traps (Figure \ref{Fig_traj_desordre}A). Therefore, the first and second trap rows are filled by the particles in the flow direction and almost all the next trap rows stay empty. Then, non-trapped particles display zigzag-like patterns around the traps, see Figure \ref{Fig_traj_desordre}A. Breaking the upstream/downstream flow symmetry thanks to the addition of disorder, modify the symmetric streamline patterns and allows a spanwise mass transport up to eventually filling a next trap. For optimized oblique MTAs, this spanwise mass transport is provided by the shift between the inlet and the outlet channels. This breaking of upstream/downstream streamlines in disordered MTA is clearly visible in the zooms of (Figure \ref{Fig_traj_desordre}B and C). Indeed, a saddle point appears on the upstream side of the trap, and recombine downstream. By breaking the symmetry, the streamlines may explore traps in the spanwise direction, a possibility that would not have been possible for a symmetric trap array. Therefore, we think that spanwise dispersion is the key mecanism allowing flow symmetry breaking and then particle trapping. Such phenomenon should be quantitatively investigated which is the subject of the next section.

\noindent 

\subsection{How spanwise dispersion influences the capture \& filling efficiencies}
\label{sec_dispersion}

\begin{figure}[H]
    \centering
    \includegraphics[width=\textwidth]{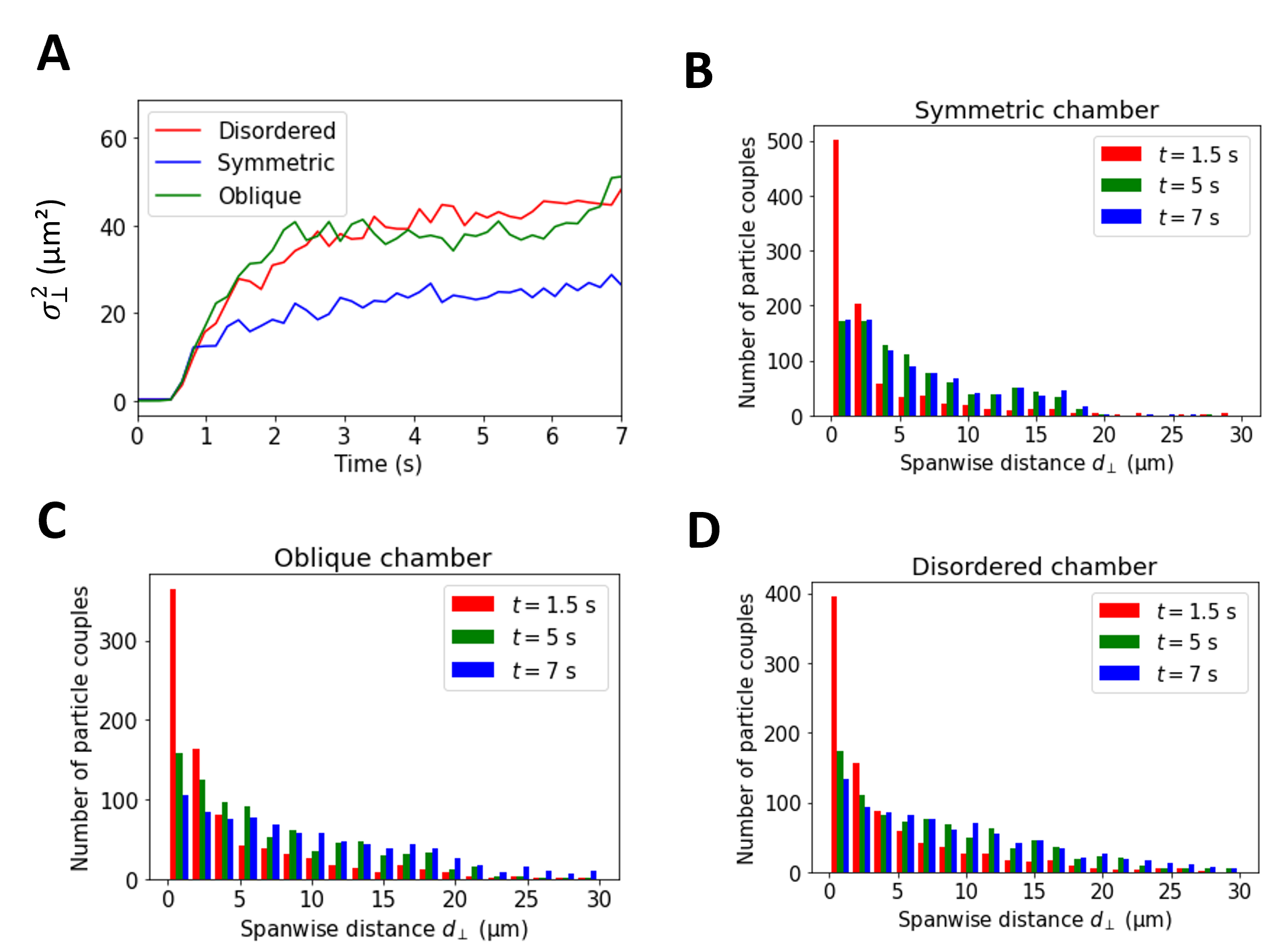}
    \caption{Dispersion regime in the three tested chambers for a $800$ traps MTA (\textbf{A}). Blue curve for the symmetric chamber, green for the optimized oblique and red for the disordered chamber.}
    \label{fig_dispersion}
\end{figure}

The particle dispersion is quantified by measuring the distance $d$ between two initially close particles at the inlet. The statistics is made on $2 \ N_{traps} -1 $ particle couples. More precisely, as we discuss the spanwise dispersion to be responsible for a better homogeneous distribution of trapped particles in the array, we measure the spanwise $d_\perp$ (along $(Oy$)) distances, see Figure \ref{fig_geom} and the corresponding mean square spanwise distance which is the usual metric to characterize dispersion:  $\sigma_\perp ^2 = \langle d_\perp^2 \rangle- \langle d_\perp \rangle^2 $. We study the particles spanwise dispersion in three cases with $800$ traps MTAs having an homothetic cavity than the previous ones : symmetric chamber (with wide inlet/outlet channels), Monte Carlo optimized oblique chamber and disordered chamber starting with a symmetric cavity ($D_f = 0,40$ and uniform distribution with wide inlet/outlet channels). The results are plotted on Figure \ref{fig_dispersion}. 

The values of the spanwise mean square distance (Figure \ref{fig_dispersion}A) evolve in a comprehensive way with respect to filling efficiency : interestingly, it seems to reach an asymptotic value for the symmetric chamber at the larger times while the values continue to increase for the disordered and oblique chambers. More precisely, this asymptote is characterized by a steady histogram of spanwise distance between couples of particles (after $t= 5$ s, Figure \ref{fig_dispersion}B) while the same histograms for oblique (Figure \ref{fig_dispersion}C) and disordered chambers (Figure \ref{fig_dispersion}D) continue to spread with time. The absence of saturation for the disordered case is in agreement with our expectation: disorder continuously disperses particles in the spanwise direction which explains why no empty region appears in the disordered chamber (see Figure \ref{fig_dispersion}B and C). We believe this metric, $\sigma_\perp^2$, is a reliable quantity to anticipate capture and filling efficiencies. 

\subsection{Experimental verification of disorder efficiency} 
\begin{figure}[H]
    \centering
    \includegraphics[width=\textwidth]{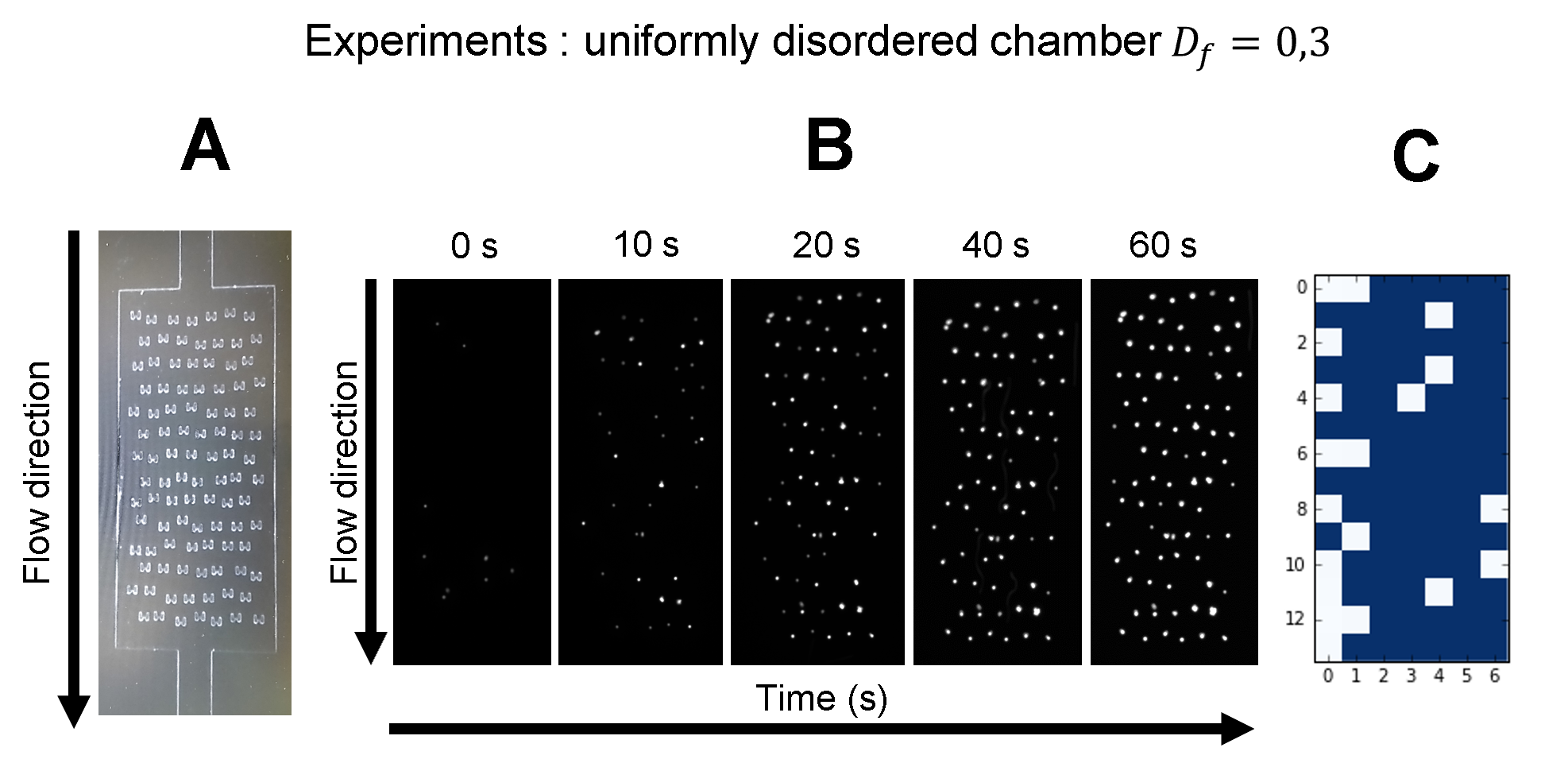}
    \caption{Trapping experiment results in disordered MTA. \textbf{A}: PDMS fabricated chamber following standard photolithography process. \textbf{B}: fluorescence microscopy images of the MTA progressive loading with $5 \ \mu$m polystyrene beads. \textbf{C}: filling matrix of the experiment : white colour for empty traps, blue for occupied traps.}
    \label{fig_exp_desordre}
\end{figure}

Since the efficiency of oblique flows was already proved experimentally with PS beads, droplets \cite{Mesdjian2021} and utilized for biological applications \cite{Pineau2022a}, we focus on the disordered chamber. The associated trapping experiments were conducted in a uniformly disordered chamber with $D_f = 0,3$. At this value, the saturation of capture efficiency is reached (Figure \ref{Fig_graph_desordre}A). The fabricated chamber is shown Figure \ref{fig_exp_desordre}A. Figure \ref{fig_exp_desordre}B shows the progressive loading of the experimental MTA, a stationary filling is reached in $60 $ s of loading. From these images and ImageJ \textregistered, we create a binar matrix of the MTA which allows to easily calculate the experimental filling efficiency (Figure \ref{fig_exp_desordre}C). As expected, the filling efficiency is high : $(E_{fill})_{exp} = 81\%$  and is close the predicted one $E_{fill} = 80\%$ of the best uniformly disordered simulated geometry, and legitimated our approach.

\section{Conclusion}
\label{section_conclusion}

In this study, we investigated the  trapping efficiency in dense MTAs. We adapted our previously presented CFD/particle model to parametrically optimise capture and filling efficiencies with five dimensionless geometric parameters. After confirming that each parameter influences the capture and filling efficiencies, we used different algorithmic optimisation approaches, named local and global. The results show the importance of creating an oblique flow to increase capture or filling efficiency. These oblique flows break the upstream/downstream symmetry of the traps and thus enhance transverse dispersion of the particles. We show that this symmetry breaking is accompanied by an increase in transverse dispersion, favouring particles to be explored over a wider area. While improving transverse dispersion is a critical point in favouring trap filling, we have shown that it is also possible to introduce disorder into the trap paving. The advantage of this configuration is that there does not appear to be any saturation of the filling. These optimisation results obtained using the numerical tool are in agreement with experimental results, demonstrating the relevance of the approach.

\section{Data availability} Data are available upon request.

\section{Acknowledgements}
This work has received support from the administrative and technological staff of “Institut Pierre-Gilles de Gennes” (Laboratoire d’excellence : ANR-$10$-LABX-$31$, “Investissements d’avenir” : ANR-$10$-IDEX-$0001$-$02$ PSL and Equipement d’excellence : ANR-$10$-EQPX-$34$).\\
NR acknowledges funding from the École Normale Supérieure de Rennes (ENS Rennes, Contrat Doctoral Spécifique Normalien) for PhD scholarship.

\section{Author contributions}
\label{section_contributions}
Concepts were proposed by MCJ and JF. Data curation was done by NR and formal analysis by MCJ and NR. Simulations were performed by NR, experiments were conducted by GF (fabrication, trapping) and NR (trapping). NR, MCJ and JF built the methodology and interpreted the results. The manuscript was written by NR and reviewed for scientific and technical aspects by MCJ and JF and for formal aspects by MCJ and RA. Work were supervised by MCJ, JF and RA. Access to experimental material was supported by JF and simulations by RA. 

\section{Conflict of interest}
There are no conflicts of interest to declare

\bibliographystyle{cas-model2-names}
\bibliography{article.bib}

\section{Supplementary information}
\label{section_SI}
\subsection{Model parameters values}
\begin{table}[H]
\begin{footnotesize}
     \begin{center}
     {\setlength\arrayrulewidth{1pt}     \begin{tabular}{c c c}
      \hline
      Fixed geometric parameters & Description & Value\\
      \hline\\
      $h$ & chamber's out-of-plane depth  & $14 \ \mu m$ \\
      $w_o$ & Traps small opening width & $4\ \mu m$\\  
      $w_{pil}$ & Trap pillar width & $7\ \mu m$\\
      $w_{sr}$ & Traps small rectangle width & $10\ \mu m$\\  
      $w_{trap}$ & Traps width & $30\ \mu m$\\ 
      $\Delta x$ & Horizontal distance between traps  & $ 50 \ \mu m $\\  
      $\Delta y$ & Vertical distance between traps & $30 \ \mu m$\\ 
      $l_{sr}$ & Trap small rectangle's length & $5 \ \mu m$\\  
      $l_{trap}$ & Trap's length & $27 \ \mu m$\\ 
      $N_{traps}$ & Number of traps in the MTA & $ 96-98 $\\\\
      \hline
      Physical parameters & Description & Values \\
      \hline\\
      $\mu $ & Fluid dynamic viscosity & $0,001 \ Pa \ s $ \\  
      $N_{part}$ & Number of particles & $2 \ N_{traps}$\\  
      $p_{in}$ & Uniform inlet pressure  & $100 \ Pa$\\  
      $p_{out}$ & Uniform outlet pressure  & $0 \ Pa$\\  
      $\rho$ & Fluid density & $1000\ kg \ m^{-3}$ \\ 
      $R_p$  & Particles radius & $2,5 \ \mu m$ \\
      $T_e$ & Study time & $3 s-32 s  $\\\\ 
      \hline
      Parametric \& optimization studies & Description & Values \\
      \hline\\
      $C$ &  Inlet/outlet centering & $ [0,100] \ \%$ \\
      $W_r$ &  Width ratio   & $ [55,97] \  \%$ \\
      $L_r$ & Length ratio  & $ [55,97] \  \%$ \\
      $A_r$ & Aspect ratio  & $0.3$,$0.4$,$0.5$,$0.7$,$1$,$1.5$,$2$,$2.7$,$3.8$ \\
      $Wch_{r}$ & Channel to chamber width ratio & $ [10,100] \ \%$ \\\\
      \hline \\
      \end{tabular}\\
      }
      \caption{The model parameters. For single parametric studies, $10$ parameters values are chosen to uniformly vary in the specified interval excepted for the $A_r$ parameter where $9$ values are chosen to have the same total number of traps $\pm 3 $ traps.}
      \label{tableau_all_param}
      \end{center}
      \end{footnotesize}
\end{table}

\subsection{Disordered traps arrays can be expanded in the spanwise direction}
Disordered MTAs generate spanwise dispersion of particles due to the trap relative positions. This characteristic offers more possibilities of the inlet/outlet channels and cavity designs than an optimized oblique chamber where all the five dimensionless parameters are fixed. As an example, we show here that when using the largest possible inlet/outlet channels ($Wch_r = 100 \% $) and expending the MTA size in the spanwise direction, the filling and capture efficiencies are maintained to their optimal values even though the MTA aspect ratio $A_r$ is modified. 

\begin{figure}[H]
    \centering
    \includegraphics[width=\textwidth]{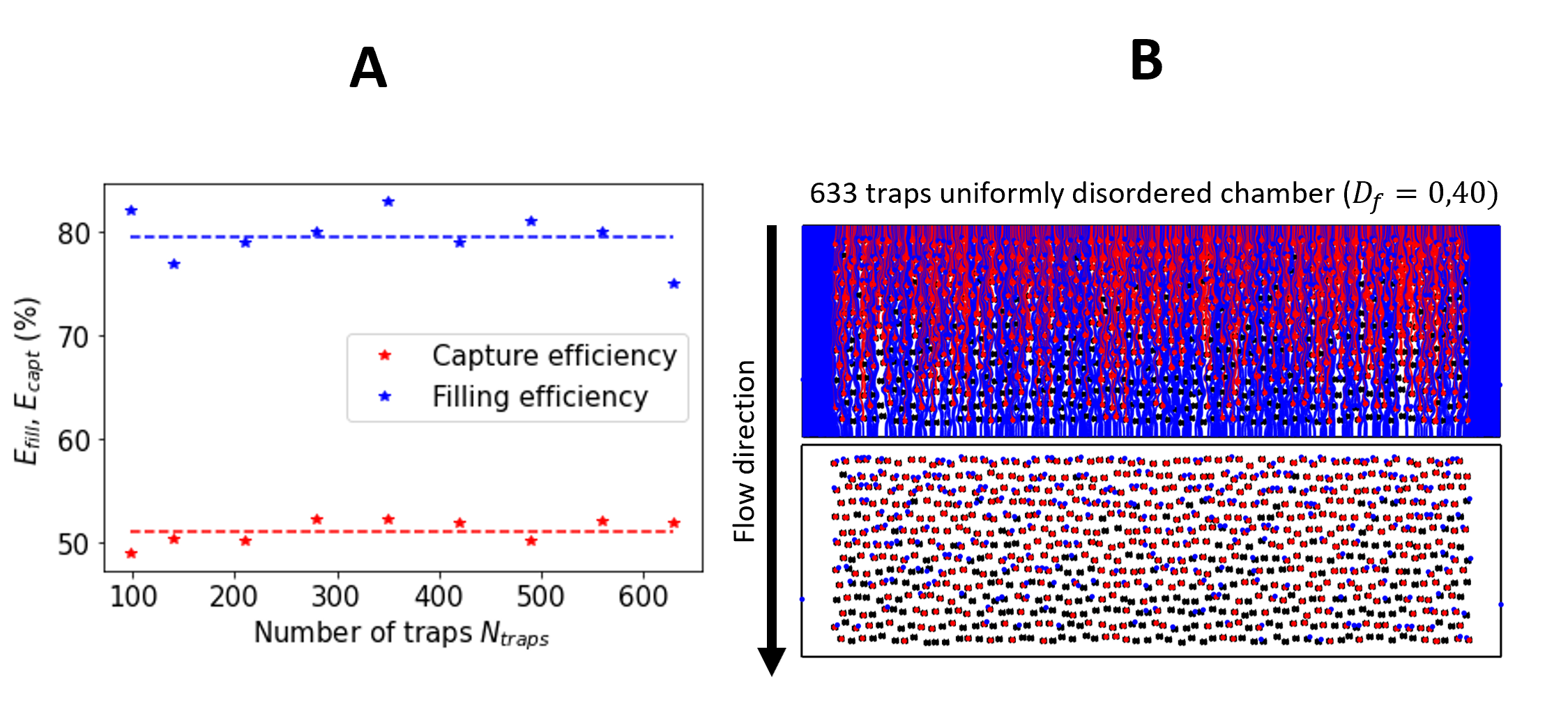}
    \caption{Evolution of capture and filling efficiencies with respect to the number of traps when expanding a $D_f = 0,40$ uniformly disordered MTA (\textbf{A}). Particles trajectories in the biggest disordered MTA simulated (\textbf{B}). Trapped particles are represented in red, the others in blue.}
    \label{fig_taille_reseau}
\end{figure}

\end{document}


\maketitle
\thispagestyle{fancy}

\begin{corrauthor}
utada.andrew.gm\at u.tsukuba.ac.jp , jacques.fattaccioli\at ens.psl.eu, jean-francois.rupprecht\at univ-amu.fr
\end{corrauthor}


\tableofcontents

\clearpage
\parindent=0mm
\section{Bacterial methods}

\vfill

Version : {\today} at \currenttime \;(UTC)